\documentstyle[aps,prl,amsmath,amssymb,psfig]{revtex}
\begin{document}

\wideabs{

\title{Mesoscopic fluctuations in superconducting dots at finite temperatures}

\author{G. Falci,$^{1}$ A. Fubini,$^{2,3}$ and A.~Mastellone$^{1}$}

\address{
  $^1$ NEST-INFM \& Dipartimento di Metodologie Fisiche e Chimiche (DMFCI),
  Universit\'a di Catania, viale A. Doria~6, 95125 Catania, Italy\\
  $^2$Dipartimento di Fisica, Universit\`a di Firenze,
  and INFM, UdR Firenze, Via G.~Sansone~1, I-50019 Sesto F.no (Fi), Italy\\
  $^3$Institut f\"ur Theoretische Physik II, Universit\"at Stuttgart,
  Pfaffenwaldring 57, 70550 Stuttgart, Germany. }
\date{\today}
\maketitle

\begin{abstract}
We study the thermodynamics of ultrasmall metallic grains with the mean level
spacing $\delta$ comparable or larger than the pairing correlation
energy in the whole range of temperatures.
A complete picture of the thermodynamics in such systems is given 
taking into account the effects of disorder, parity and classical and quantum
fluctuations. Both spin susceptibility and specific heat turn out to be
sensitive probes to detect superconducting correlations in such
samples.
\end{abstract}

}

Superconducting coherence is expected to be completely washed out in
small metallic grains, where  the 
average level spacing $\delta \sim 1/({\cal N}(0)V)$ may be 
comparable or even larger than the BCS energy scale $\Delta$~\cite{kn:early}. 
Despite of that, a series of experiments~\cite{RBT,vonDelft} by  
Ralph, Black and Tinkham (RBT) revealed a 
gap related to superconductivity in individual nanosized metallic 
grains, $\delta \lesssim \Delta$, with large electrostatic energy
fixing the total number $N$ of electrons. The BCS description of 
superconducting coherence is no longer valid and
the characterization of ``superconductivity'' from the bulk to the 
ultrasmall-grain regime has been studied in recent theoretical 
works~\cite{Matveev97,Mastellone98,Berger98,kn:various}.
Starting from the standard pairing Hamiltonian~\cite{vonDelft}, 
several spectral features at $T=0$ were calculated, which turn out to be
universal functions of 
the single scaling parameter $\delta/\Delta$~\cite{Mastellone98}.
It is remarkable that pairing determines strong fluctuational
``superconductivity'' even in ultrasmall grains~\cite{Matveev97}, 
where $\delta \gtrsim \Delta$ (we call this the 
``dot'' regime).  
As in micron-size samples~\cite{parity}, parity effects are present, i.e. 
physical properties depend on $N$ being odd or even. An exact solution for 
the pairing problem in finite systems 
was found long ago by Richardson and Sherman (RS)~\cite{RICHARDSON}.

\begin{figure}
\centerline{\psfig{figure=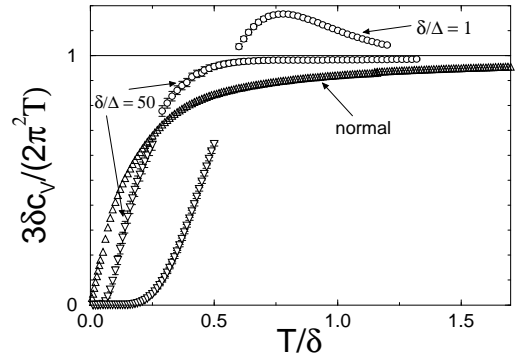,width=65mm,angle=0}}
\caption{Rescaled specific heat $ c_{Ve} \delta / \gamma T$
vs. temperature in units of $\delta$, disordered case. Circles: the
RPA$^\prime$ results; triangles: the exact canonical results.}
\label{fig:specific-heatGOE}
\end{figure}

\begin{figure}
\centerline{\psfig{figure=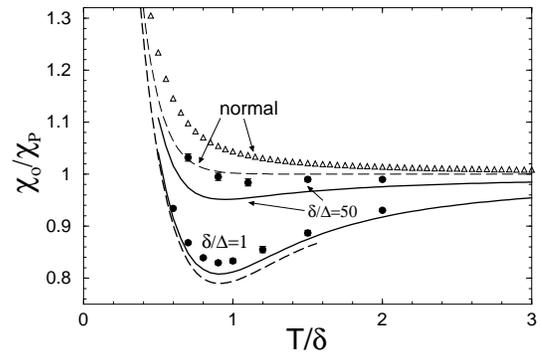,width=70mm,angle=0}}
\caption{Spin susceptibility for dots with odd $N$. 
Equally spaced noninteracting
single-particle levels case as given by the RS canonical 
solution (dashed lines), and the RPA$^\prime$ (solid), are compared 
with results for GOE
distributed single-particle levels obtained by RPA$^\prime$ (circles)
and with the exact canonical calculation (triangles).}
\label{fig:chi-odd-disordered}
\end{figure}

In this work the thermodynamics 
of an {\em ensemble} of monodispersed superconducting dots is studied. 
The specific heat $c_V(T)$ 
(see Fig.~\ref{fig:specific-heatGOE}) and the spin susceptibility 
$\chi(T)$ (see Fig.~\ref{fig:chi-odd-disordered})
are obtained in the whole temperature range. 
Our work is motivated by the fact that thermodynamic properties 
are a unique  experimental tool in detecting unambiguous traces of 
superconducting correlations in the dot regime, where  
tunneling spectroscopy on {\em individual} dots
is not sensitive enough~\cite{RBT}.
Di Lorenzo et al.~\cite{DiLorenzo99} proposed that a possible 
signature of pairing for $\delta \gtrsim \Delta$, 
is the reentrant behavior of the spin 
susceptibility $\chi_o(T)$ in dots with odd $N$, 
which is due to the combined parity  and interaction effects,
using a simple equally spaced spectrum for the single particle energies.

In ensembles of grains the single-particle spectrum is statistically 
distributed and characterized by level repulsion~\cite{kn:metha}. 
The statistics has a universal character due to general symmetry 
properties and is described by Random Matrix Theory (RMT). The 
thermodynamics of ensembles of normal metal grains was  
studied long ago~\cite{denton}, but only
in recent years the first experimental evidence of mesoscopic 
fluctuations and level repulsion in normal metal grains was 
provided with the observation of the $T^2$-dependence of the specific 
heat and of other thermodynamic quantities~\cite{kn:qse-exp}.
Pioneering experimental works on thermal properties of small
superconducting grains date back to the 80s \cite{shapira}.

Concerning superconductors, level statistics 
was found to yield enhanced superconductivity and parity 
effects in micron-size samples~\cite{kn:smith-ambegaokar-96}. 
Similar results have been recently found also in the 
dot regime~\cite{Sierra99} at $T=0$. 
Here we focus on ensembles of dots at finite temperature. 
They exhibit detectable features characteristic of the interplay between
pairing interaction and the universal statistics of mesoscopic
fluctuations, which determines the physics.

Our starting point is the Hamiltonian 
\begin{equation}
        {\cal H} = \sum_{{\alpha, \sigma}}
          (\epsilon_\alpha \!- \sigma \mu_B H)\,
        c_{\alpha,\sigma}^{\dagger}  c_{\alpha,\sigma}\! - \lambda \;
        T^\dagger \, T ~,
\label{eq:BCS-pairing-hamiltonian-H}
\end{equation}
where $\alpha$ spans a shell of $\Omega$ pairs ($\sigma=\pm$) of
single particle energy levels with statistically distributed
energies $\epsilon_\alpha$ and annihilation
operator $c_{\alpha,\sigma}$.  The magnetic field $H$ enters in
the Zeeman form ($\mu_B$ is the Bohr magneton).  For
$H=0$ the pair of states $\sigma=\pm$ are degenerate and
time-reversed. 
The operator $T = \sum_{\alpha=1}^{\Omega}c_{\alpha,-}c_{\alpha,+}$ 
appears in the interaction term, which scatters pairs with amplitude $\lambda$, 
so a pair of levels $\sigma=\pm$ occupied by a single electron 
is blocked~\cite{vonDelft}. Relevant energy scales are the average 
level spacing $\delta$ and the BCS energy, defined as
$\Delta = \Omega \delta / [2 \sinh(\delta/\lambda)]\,$.
This model stems from a rather general description of electron-electron 
interaction in the metallic regime
(large dimensionless conductance, $g=E_T/\delta\gg 1$, where
$E_T$ is the Thouless energy) which has been recently 
proposed~\cite{altshuler}. 
The Hamiltonian Eq.(\ref{eq:BCS-pairing-hamiltonian-H}) describes 
universal properties of pairing
interaction in disordered metallic dots, 
with fixed $N$ and negligible exchange interaction.
We assume that 
the energies $\epsilon_\alpha$ are distributed according to the  Gaussian
Orthogonal Ensemble (GOE), which describes mesoscopic fluctuations
when no spin-orbit interaction is present and time-reversal symmetry
is preserved\cite{kn:metha}. 

Level statistics has a drastic effect if $T \ll \delta$,
where the thermodynamics is determined by samples with
$\delta_1$, the closest level spacing to the Fermi energy, much smaller than
the average $\delta$. The relevant excitations involve only few levels around 
the Fermi energy, and excitation energies can be estimated 
by solving the problem in a small shell, $\Omega^{\prime} \ll \Omega$, 
with renormalized coupling constant 
$\lambda \equiv \tilde{\lambda}_{\Omega} \to 
\tilde{\lambda}_{\Omega^{\prime}}$~\cite{Matveev97,Berger98,DiLorenzo99}.
Let's consider dots with even $N$ and estimate the leading low 
temperature contribution to $c_{Ve}(T)$. 
In the regime 
$\delta \gg \Delta$, the system is first renormalized to an effective one with 
two electrons in two doubly degenerate levels, and 
interaction $\tilde{\lambda}_2$. 
We have to consider six states, and the relative excitation energies:
ground state, excited state of paired electrons 
($\tilde{E}_{\rm p}(\delta_1)= 2 (\delta_1^2+\tilde{\lambda}_2^2)^{1/2}$)
and four states corresponding to a broken pair 
with different spin configurations 
($\tilde{E}_{\rm bp}(\delta_1)=
\tilde{\lambda}_2 + \tilde{E}_{\rm p}(\delta_1)/2$). 
Notice that even if the coupling is weak, $\tilde{\lambda}_2 \ll \delta$, 
dominant configurations have $\delta_1 \lesssim \tilde{\lambda}_2\,$. 
The response of a set of grains is evaluated by retaining 
only fluctuations of $\delta_1$, governed by the level spacing 
distribution  $P_1^{{}^{\rm GOE}}(\delta_1)$. In this way we obtain the 
analytic form for $T \ll \delta$.
The leading term for $T \ll \tilde{\lambda}_2$ reads 
\begin{eqnarray}
\label{eq:cV-even-low-T}
c_{Ve}(T) \,=\, 3 \, \pi^2 \; { \tilde{\lambda}_2^3 \over \delta^2 T} 
\; {\rm e}^{-{ 2 \tilde{\lambda}_2 \over T}}
\end{eqnarray}
and shows a gap $2  \tilde{\lambda}_2$.  
In the limit $\tilde{\lambda}_2 \ll T \ll \delta$ the full analytic result 
reproduces the known result for normal metal grains~\cite{denton}. 
For $c_{Vo}(T)$ we have to consider an effective 
three (fluctuating) level problem, and we obtain 
\begin{equation}
\label{eq:cV-odd-low-T}
c_{Vo}(T) \;=\; \frac{3}{2}\pi^2 \zeta(3) \bigg(\frac{T}{\delta}\bigg)^2 +
O\bigg(\frac{\tilde{\lambda}_3}{\delta}\bigg)^2~, 
\end{equation}
which shows no trace of a gap. 
The contribution of superconducting
correlations for $T\ll\tilde{\lambda}_3$ is a correction of 
the result for normal grains~\cite{denton}, because the most important 
low-energy excitations are obtained by moving the unpaired electron.

Dots at intermediate temperatures can be studied numerically by the massive 
use of the RS exact solution.
Noninteracting random 
spectra are obtained by diagonalizing $500\times 500$
random matrices belonging to the
GOE and taking the central levels of the
resulting semicircular distribution, in order to prevent undesired
border effects. For each disorder realization we evaluate the partition function
from the universal Hamiltonian Eq.(\ref{eq:BCS-pairing-hamiltonian-H})
using the RS solution and calculate the average free energy.
For the curve $\delta/\Delta = 50$ shown in Fig.~\ref{fig:specific-heatGOE} we
used sets of 50 levels. At intermediate temperatures ($T
\lesssim 0.25 \delta$ for the curve in
Fig.~\ref{fig:specific-heatGOE}) we used 500 realizations
and an energy cutoff $5 \delta$. 
At larger coupling $\lambda$, corresponding to the crossover region
$\delta/\Delta \gtrsim 1$, and when excited states with random 
spectra are concerned, singularities in the RS equations 
become untractable. Then the low-temperature behavior for $\delta/\Delta
\gtrsim 1$ shown in Fig.~\ref{fig:specific-heatGOE} was studied 
following Ref.~\cite{Berger98}: for each 
realization of disorder and configuration of blocked levels 
the interaction is scaled down to some
low-energy cutoff and the effective small ($\le 12$ levels) system,
and is then diagonalized.  

Both methods described above cannot be used to
obtain results at temperatures $T \sim \delta,\Delta$ where $c_V(T)$ is 
expected to show anomalies. In this regime a huge
number of states (at least $5\cdot10^4$ already at $T\sim\delta$) are needed
for {\em each} single disorder realization. 
Then for 
high temperature we use a functional technique in the parity-projected
(PP) grand canonical ensemble. Following Janko et al.~\cite{parity} we
introduce the even and odd $N$ partition function 
\begin{equation}
\label{e.Zeo}
        Z_{e/o}(T,\mu) = \frac12 \sum_{N=0}^{\infty} e^{ \mu N/T}
        [1 \pm e^{i \pi N}] Z(T,N) ~,
\end{equation}
for each given realization of disorder.
Here $Z(T,N)$ is the canonical partition function so $Z_{e/o}(T,\mu)$
involve only sectors where $N$ is not fixed but its parity is. 
The effective imaginary-time
action is obtained from Eq.(\ref{eq:BCS-pairing-hamiltonian-H}) in the
standard way and $Z_{e/o}(T,\mu)$ can be expressed 
as a path integrals over a Hubbard-Stratonovic (HS) auxiliary field 
$\Delta(\tau)$~\cite{DiLorenzo99}.  
This non-linear functional integral is evaluated in the 
RPA$^\prime$, an approximation which has been
widely employed 
to study the Anderson model\cite{kn:wang69}.
We extract the static part of the HS field in the Matsubara-Fourier 
representation,
$\Delta(\tau) = \Delta_0 +\sum_{n\neq 0} \Delta(\omega_n)
\exp(-i\omega_n \tau)\,$, evaluate quantum corrections by including the 
contribution of small-amplitude deviations around a generic 
static path $\Delta_0$ and perform numerically the remaining
ordinary integral over $\Delta_0$. We used systems of 100 levels distributed according to GOE. Finally the resulting free energy
is averaged over 400 samples.  
Results for $c_{Ve}(T)$ are shown in 
Fig.\ref{fig:specific-heatGOE}\,.
For large enough - albeit still nanosized - grains a 
finite temperature anomalous enhancement develops, which 
can be attributed to 
the presence of pairing interactions.

We now study the spin susceptibility. Using the same arguments leading to 
Eqs.(\ref{eq:cV-even-low-T},\ref{eq:cV-odd-low-T}) we can conclude that at 
low temperatures the susceptibility of grains with even $N$  
is exponentially suppressed, 
so we concentrate on $\chi_o(T)\,$. For equally spaced 
single particle spectrum $\chi_o(T)\,$ shows a reentrant behavior for 
non-vanishing pairing interaction, because at low temperature the leading 
contribution is the $1/T$ Curie-like coming from the unpaired electron,
whereas the response of the paired electrons exponentially increases with 
$T$ and becomes dominant at larger temperatures~\cite{DiLorenzo99}. 
As compared with results for equally spaced spectra, we notice that
level statistics triggers two competing 
effects:  the 
availability of low-lying excitations tends to increase 
$\chi_o(T)\,$, which would wash out the reentrance; on the other hand 
disorder enforces superconductivity, which would determine the opposite 
trend. 

We evaluate $\chi_o(T)\,$ using the PP partition 
function Eq.(\ref{e.Zeo}) in the RPA$^\prime$.
Results (Fig.~\ref{fig:chi-odd-disordered}) 
clearly show that $\chi_o(T)$ is reentrant
also in disordered samples. 
The reentrance is slightly less pronounced 
than for regular spectra and it is present always but for 
unphysically small grains. 
Notice that the ``superconducting'' $\chi_o(T)$ is well suppressed  
with respect to normal metal grains, confirming the 
expectation that disorder favors superconductivity. 
The leading behavior of $\chi_o(T)$ for $T \ll \delta$ is unaffected by 
level statistics.

We now discuss experimental signatures of superconducting correlations. 
The main point is that in ensembles of dots available in experiments 
roughly 50\% have odd $N$ and 50\% have even $N$. 
At low temperatures the effect of pairing is detectable in the total 
$c_V(T)$: in an ensemble of normal metal grains the 
relative contributions to the ensemble specific-heat of dots with even
$N$ is about $63\%$~\cite{denton}. Pairing suppresses this contribution,  
Eq.(\ref{eq:cV-even-low-T}), whereas in practice it does not affect  
$c_{Vo}(T)$,  Eq.(\ref{eq:cV-odd-low-T}). Thus pairing reduces the total
$c_V(T)$ of $\sim\,2/3$. Another signature is that 
$c_{V}(T)$ increases with an applied magnetic field. 
Other signatures can be found at intermediate $T$:
The total susceptibility $\chi(T)$ is expected
to show reentrant behavior ($\chi(T) \approx \chi_o(T)$, since 
$\chi_e(T)$ is exponentially suppressed), except for extremely weak 
interactions or extremely small grains. Another signature is 
the {\em increase} of $\chi(T)$ with the applied magnetic field.
In the same temperature regime ensembles of dots with 
$\delta/\Delta \sim 1$ show an anomalous enhancement arising 
from the contribution of $c_{Ve}(T)$ which can be shifted towards 
lower temperatures and the progressively suppressed by applying a 
magnetic field. The whole phenomenology could provide unambiguous evidence 
of the presence of pairing correlations even in the dot regime.

The procedure used to obtain the analytic results for $T \ll \delta$,
neglects fluctuations of the renormalized
coupling $\tilde{\lambda}_2$ ($\tilde{\lambda}_3$ for dots with odd $N$).
They arise from fluctuations of larger shells of levels which may be
systematically accounted for. In normal metal grains~\cite{denton}, they 
can be neglected for $T \ll \delta$.  
We checked numerically that $c_{Ve}$ is exponentially suppressed and that 
for $\delta/\Delta \gg 1$ the exponent roughly agrees with the scaling form 
discussed in Ref.~\cite{Matveev97}. 
We used a large number of realizations ($\sim
10^4-10^5$) since at very low $T$, 
the system is very sensitive to the statistic. Detailed account will be 
given elsewhere.

Some remarks on the results presented above are clearly stated if we 
draw a comparison with a dot having equally spaced noninteracting levels, 
$\epsilon_\alpha=\alpha \delta$. In this case 
all the thermodynamic quantities show gapped behavior 
for $T \ll \delta$. This is  
essentially due to the finite level spacing~\cite{denton,DiLorenzo99}, 
for instance the even specific heat has
a gap $\delta + \tilde{\lambda}_2\,$. 
On the contrary, in ensembles of grains gapped behavior is not the rule. 
When a gap is present, it is due {\em only} to the pairing interaction, as in 
the case of $c_{Ve}(T)$, Eq.(\ref{eq:cV-even-low-T}).

\begin{figure}
\centerline{\psfig{figure=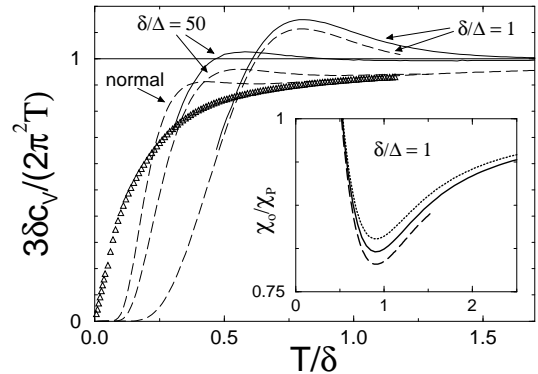,width=69mm,angle=0}}
\caption{Thermodynamic quantities for dots with equally spaced spectra, 
calculated by the RS solution (dashed lines),
the SPA (dotted) and the RPA$^\prime$ (solid) as a function of 
temperature in units of $\delta$. Main figure: the rescaled specific
heat $c_{Ve}/ \gamma T$.
The curve for normal metal disordered dots is plotted for comparison
(triangles), see text. In the inset the odd
spin susceptibility $\chi_o(T)$, see text.}
\label{fig:specific-heat}
\end{figure}

In Fig.(\ref{fig:specific-heat}) we present results at finite temperatures
for $c_{Ve}(T)$ and $\chi_o(T)$ for regular spectra, 
$\epsilon_\alpha = \alpha\delta$. 
In this case the exact RS solution 
allows to reach temperatures slightly larger than $\delta$; 
larger temperatures ($T \gtrsim \delta$) were studied by 
the PP approach, Eq.~(\ref{e.Zeo}).  
In the inset we show $\chi_o(T)$, calculated using the canonical RS solution,
and the PP grand canonical approach in the RPA$^\prime$ and in the 
Static Path Approximation (SPA)~\cite{DiLorenzo99}. In the SPA 
fluctuations around the static path $\Delta_0$ are neglected 
when evaluating the PP partition function Eq.~(\ref{e.Zeo}). The SPA accounts
for thermal fluctuations and it has been successfully employed to study the
thermodynamics of micrometer size grains~\cite{Muelschlegel2}. 
By comparing the SPA with the RPA$^\prime$ results we can quantify quantum 
corrections. As expected quantum superconducting fluctuations tend to enhance the
features of pairing: the reentrance in $\chi_o(T)$ is
deepened compared to the SPA result getting closer to the exact result
for $T \sim \delta$. Differences between the RPA$^\prime$ and the
exact result reflect the different sets of excitations 
available in the canonical and in the PP grand-canonical ensembles.
The discrepancy is small and probably undetectable when pairing 
interaction is active, 
so RS solution and RPA$^\prime$ can be used in combination.
Small differences are present also in $c_{Ve}(T)$ 
(Fig.\ref{fig:specific-heat}) and
they are expected to 
persist at $T \gg \delta$. In this latter regime $c_{Ve}(T)$ approaches 
the behavior of normal metals,
where it is known that the canonical $c_V(T)$ is
smaller than the grand-canonical one, asymptotically by a quantity
$\frac12 k_{\rm B}$~\cite{Muelschlegel2}.


We now focus on the anomaly at intermediate temperatures appearing in 
$c_{Ve}(T)$. 
The qualitative behavior of $c_{Ve}(T)$ for $\delta/\Delta \sim 1$ 
is reminiscent of what observed in bulk superconductors, but this 
is partly due to the special choice of the equally spaced noninteracting 
spectrum. In fact the $T \lesssim \delta$ enhancement in the 
specific heat is present also if the pairing interaction is switched off. 
This artifact is not present for normal grains with levels 
$\epsilon_\alpha$ distributed  according to the GOE statistics 
(symbols in  Fig.\ref{fig:specific-heat})~\cite{denton}. 
Thus the analysis of the effects of level statistics is necessary for 
a reliable study of the thermodynamics of small
grains and gives an unequivocal discrimination of the effects
of the pairing interaction.

In summary we studied the thermodynamics of ensembles of ultrasmall
superconducting grains, addressing for the first time this problem in
the dot regime. We draw a picture of the physics which comes from the 
interplay of disorder, strong pairing fluctuations and 
fixed number of electrons in each dot. 
Thermodynamic properties are a unique  experimental tool in detecting 
unambiguous traces of superconducting correlations in dots. We found that
several signatures of pairing (the low and intermediate temperature 
anomalies of the specific heat and the reentrant spin susceptibility) 
are experimentally detectable in ensembles of dots, despite superconductivity 
in the BCS sense breaks down.
Finally the low-temperature behavior of $c_V(T)$ may provide direct evidence
of the effects of level statistics~\cite{kn:qse-exp} in interacting 
metallic systems.

{\em Acknowledgments.}  We acknowledge Rosario Fazio, 
A. Di Lorenzo, L. Amico, G. Giaquinta, E. Piegari,
G. Sierra, V. Tognetti.  We acknowledge support of COFIN-MURST and EU 
grant TMR-FMRX-CT-97-0143. AF acknowledges
support of ``Fondazione Della Riccia''.

\end{document}